\documentclass[journal,compsoc]{IEEEtran}
\usepackage[utf8]{inputenc}
\usepackage{amsmath}
\usepackage[noadjust]{cite}
\usepackage{todonotes}
\usepackage[binary-units]{siunitx}
\usepackage{booktabs}
\usepackage{hyperref}
\usepackage{textcomp}
\usepackage{cleveref}
\usepackage{graphicx}
\usepackage[caption=false,font=normalsize,labelfont=sf,textfont=sf]{subfig}
\usepackage{tikz}
\usepackage{scalerel}
\usepackage{pgfplots}

\Crefname{equation}{Eq.}{Eqs.}
\Crefname{figure}{Fig.}{Figs.}
\Crefname{tabular}{Tab.}{Tabs.}

\usetikzlibrary{calc,intersections,shapes.geometric,positioning}
\makeatletter
\pgfkeys{/pgf/.cd,
  cube offset x/.initial=8pt,
  cube offset y/.initial=8pt
}
\pgfdeclareshape{cube}
{
  \inheritsavedanchors[from=rectangle]
  \inheritanchorborder[from=rectangle]
  \inheritanchor[from=rectangle]{north}
  \inheritanchor[from=rectangle]{north west}
  \inheritanchor[from=rectangle]{north east}
  \inheritanchor[from=rectangle]{center}
  \inheritanchor[from=rectangle]{west}
  \inheritanchor[from=rectangle]{east}
  \inheritanchor[from=rectangle]{mid}
  \inheritanchor[from=rectangle]{mid west}
  \inheritanchor[from=rectangle]{mid east}
  \inheritanchor[from=rectangle]{base}
  \inheritanchor[from=rectangle]{base west}
  \inheritanchor[from=rectangle]{base east}
  \inheritanchor[from=rectangle]{south}
  \inheritanchor[from=rectangle]{south west}
  \inheritanchor[from=rectangle]{south east}
  \backgroundpath{
    \southwest \pgf@xa=\pgf@x \pgf@ya=\pgf@y
    \northeast \pgf@xb=\pgf@x \pgf@yb=\pgf@y
    \pgfmathsetlength\pgfutil@tempdima{\pgfkeysvalueof{/pgf/cube offset x}}
    \pgfmathsetlength\pgfutil@tempdimb{\pgfkeysvalueof{/pgf/cube offset y}}
    \def\ppd@offset{\pgfpoint{\pgfutil@tempdima}{\pgfutil@tempdimb}}
    \pgfpathmoveto{\pgfqpoint{\pgf@xa}{\pgf@ya}}
    \pgfpathlineto{\pgfqpoint{\pgf@xb}{\pgf@ya}}
    \pgfpathlineto{\pgfqpoint{\pgf@xb}{\pgf@yb}}
    \pgfpathlineto{\pgfqpoint{\pgf@xa}{\pgf@yb}}
    \pgfpathclose
    \pgfpathmoveto{\pgfqpoint{\pgf@xb}{\pgf@ya}}
    \pgfpathlineto{\pgfpointadd{\pgfpoint{\pgf@xb}{\pgf@ya}}{\ppd@offset}}
    \pgfpathlineto{\pgfpointadd{\pgfpoint{\pgf@xb}{\pgf@yb}}{\ppd@offset}}
    \pgfpathlineto{\pgfpointadd{\pgfpoint{\pgf@xa}{\pgf@yb}}{\ppd@offset}}
    \pgfpathlineto{\pgfqpoint{\pgf@xa}{\pgf@yb}}
    \pgfpathmoveto{\pgfqpoint{\pgf@xb}{\pgf@yb}}
    \pgfpathlineto{\pgfpointadd{\pgfpoint{\pgf@xb}{\pgf@yb}}{\ppd@offset}}
  }
}
\makeatother
\definecolor{clr1}{RGB}{240,240,240}
\tikzset{>=latex}
\tikzstyle{square} = [rectangle, thin, draw=black, minimum size=30,
    transform shape]
\tikzstyle{svobb} = [circle, thin, draw=black, fill=clr1, minimum size=65,
    transform shape]
\tikzstyle{svo} = [cube, thin, draw=black, fill=white, minimum size=37.5,
    anchor=south west, transform shape]

\usetikzlibrary{svg.path}

\definecolor{orcidlogocol}{HTML}{A6CE39}
\tikzset{
  orcidlogo/.pic={
    \fill[orcidlogocol] svg{M256,128c0,70.7-57.3,128-128,128C57.3,256,0,198.7,0,128C0,57.3,57.3,0,128,0C198.7,0,256,57.3,256,128z};
    \fill[white] svg{M86.3,186.2H70.9V79.1h15.4v48.4V186.2z}
                 svg{M108.9,79.1h41.6c39.6,0,57,28.3,57,53.6c0,27.5-21.5,53.6-56.8,53.6h-41.8V79.1z M124.3,172.4h24.5c34.9,0,42.9-26.5,42.9-39.7c0-21.5-13.7-39.7-43.7-39.7h-23.7V172.4z}
                 svg{M88.7,56.8c0,5.5-4.5,10.1-10.1,10.1c-5.6,0-10.1-4.6-10.1-10.1c0-5.6,4.5-10.1,10.1-10.1C84.2,46.7,88.7,51.3,88.7,56.8z};
  }
}

\newcommand\orcidicon[1]{\href{https://orcid.org/#1}{\raisebox{1pt}{\scalerel*{
\begin{tikzpicture}[yscale=-1,transform shape]
\pic{orcidlogo};
\end{tikzpicture}
}{|}}}}
\pgfplotsset{barplot/.style={ybar, point meta=y, nodes near coords, nodes near
    coords style={font=\scriptsize, rotate=90, anchor=west},align=center,
    tick label style={font=\scriptsize}, title style={font=\small},
    enlarge x limits=0.2, xtick=data, height=5cm, width=0.45\textwidth,
    ylabel style={font=\footnotesize}}}

\definecolor{matlab1}{rgb}{0, 0.4470, 0.7410}
\definecolor{matlab2}{rgb}{0.8500, 0.3250, 0.0980}
\definecolor{matlab3}{rgb}{0.9290, 0.6940, 0.1250}
\def\arxiv{}

\begin{document}

\title{Efficient Animation of Sparse Voxel Octrees\\for Real-Time Ray Tracing}

\author{%
    Asbjørn~Engmark~Espe~\orcidicon{0000-0003-1687-889X}, %~\IEEEmembership{Student~Member,~IEEE,}
    Øystein~Gjermundnes\IEEEauthorrefmark{1}~\orcidicon{0000-0002-3500-8366}, %~\IEEEmembership{Student~Member,~IEEE,}
    and Sverre~Hendseth~\orcidicon{0000-0001-7232-6868}%,~\IEEEmembership{Member,~IEEE}
\IEEEcompsocitemizethanks{%
    \IEEEcompsocthanksitem The authors are with the Department of Engineering
    Cybernetics,\protect\\
    and \IEEEauthorrefmark{1}Department of Electronic Systems,\protect\\
    Norwegian University of Science and Technology, Trondheim, Norway.
    \IEEEcompsocthanksitem
    E-mail: \{asbjorn.e.espe, oystein.gjermundnes, sverre.hendseth\}@ntnu.no
    }%
%\thanks{Manuscript received December XXth, 2019}%; revised March XXth, 2020; Accepted April XXth, 2020.}%
}

\newcommand\copyrighttext{%
  \scriptsize \textcopyright\;2019 IEEE. Personal use of this material is
  permitted. Permission from IEEE must be obtained for all other uses, in any
  current or future media, including reprinting/republishing this material for
  advertising or promotional purposes, creating new collective works, for resale
  or redistribution to servers or lists, or reuse of any copyrighted component
  of this work in other works.
  %DOI: \href{<http://tex.stackexchange.com>}{10.0001/uhtue.49}
  }
\newcommand\copyrightnotice{%
\begin{tikzpicture}[remember picture,overlay]
\node[anchor=south,yshift=10pt] at (current page.south) {\fbox{\parbox{\dimexpr\textwidth-\fboxsep-\fboxrule\relax}{\copyrighttext}}};
\end{tikzpicture}%
}

\IEEEtitleabstractindextext{%
\begin{abstract}
A considerable limitation of employing sparse voxels octrees (SVOs) as a model
format for ray tracing has been that the octree data structure is inherently
static. Due to traversal algorithms' dependence on the strict hierarchical
structure of octrees, it has been challenging to achieve real-time performance
of SVO model animation in ray tracing since the octree data structure would
typically have to be regenerated every frame. Presented in this article is a
novel method for animation of models specified on the SVO format. The method
distinguishes itself by permitting model transformations such as rotation,
translation, and anisotropic scaling, while preserving the hierarchical
structure of SVO models so that they may be efficiently traversed. Due to its
modest memory footprint and straightforward arithmetic operations, the method is
well-suited for implementation in hardware. A software ray tracing
implementation of animated SVO models demonstrates real-time performance on
current-generation desktop GPUs, and shows that the animation method does not
substantially slow down the rendering procedure compared to rendering static
SVOs.
\end{abstract}

\begin{IEEEkeywords}
Voxel, octree, animation, ray tracing, computer graphics.
\end{IEEEkeywords}}

% make the title area
\maketitle

\ifdefined\arxiv
\copyrightnotice
\fi

\IEEEdisplaynontitleabstractindextext

\IEEEraisesectionheading{\section{Introduction}}
\IEEEPARstart{T}{he} overarching goal of computer graphics is to use a computer
to deterministically render images based on a specification of some form. These
images may be stored for later consumption, or they may be presented in
real-time on a display as part of a graphics pipeline. In the last decades,
rasterisation has emerged as the most popular technique for rendering real-time
computer graphics. However, recent developments have shown that ray tracing may
be a legitimate alternative to rasterisation for certain real-time rendering
applications \cite{caulfield18}. There are indications of a new paradigm being
established in computer graphics, the central tenet of which is that ray tracing
may be used in conjunction with traditional rasterisation rendering, so that
each technique is utilised towards its strengths. Nvidia demonstrated this in
the autumn of 2018 with their introduction of a new range of consumer graphics
processing units (GPUs) containing dedicated hardware for acceleration of ray
tracing \cite{nvidia_rtx}.

A common application of ray tracing is the rendering of three-dimensional
volumetric models. The sparse voxel octree (SVO) is a specific flavour of the
octree data structure that may be used to store such volumetric model data.
SVOs are especially attractive for ray tracing since they have been studied
thoroughly, resulting in a vast body of published research on their optimised
traversal as part of a ray tracing pipeline. Hence, if a three-dimensional
object can be adequately represented by an SVO model, it may be rendered
efficiently and with sufficient visual fidelity through ray tracing. The
structure of an SVO closely resembles that of the traditional sparse octree,
but differs in that instead of using the data structure to subdivide space or
sort objects, the octree itself directly encodes the volumetric data. Certain
properties are associated with the tree nodes, which enable the data structure
to natively describe arbitrary objects as sets of voxels. The model fidelity is
consequently limited only by the hierarchical depth of the octree. The challenge
of using SVOs for real-time computer graphics is that efficient animation is
hindered by their strict hierarchical structure. In an arbitrary animation
sequence, each node of the octree may be altered, which would typically require
that the tree structure be regenerated every frame.

The purpose of this research is to derive a new method that efficiently
facilitates the animation of otherwise static SVO models. The method supports a
certain subset of animation transformations which will collectively be referred
to as rigid-body animation. As defined in this work, rigid-body animation
comprises all transformations that do not alter the internal data of the model.
This means, in turn, that any transformation that is to be applied to the model
must be applied equally to all the model's internal vertices, and that during
the animation sequence, the relative positions and orientations of all nodes
will remain unchanged in model space.

In addition to the animation method itself---which will be introduced in
\Cref{sec:solution}---a set of optimisation techniques will be presented and
discussed in \Cref{sec:optimisations}. These techniques are included since they
are relevant for future use of the animation method, and will generally benefit
an implementation in terms of performance. They were employed to great success
in the authors' own software implementation, which will be evaluated alongside
the animation method in \Cref{sec:evaluation}.

\section{Related Works}\label{sec:related_works}
No general technique for animation of SVO models optimised for ray tracing was
found in the literature. Nonetheless, a single attempt was found for a
special-case form of animation of SVO models. \cite{bautembach11} introduces a
method for SVO animation based on the idea that each leaf node of the tree is an
individual \textit{atom} that may be animated. However, the method is not
applicable for ray tracing. As is explicitly stated by the work's author, Dennis
Bautembach, the hierarchical structure of the SVO model is destroyed as part of
the animation process. Consequently, most ray tracing algorithms will no longer
work, since efficient intersection tests are effectively prohibited. Bautembach
therefore resorts to rasterisation in order to render the animated sparse voxel
octrees. Thus, to the best of the authors' knowledge, the method presented in
this article is the only efficient technique for the animation of SVOs for use
in ray tracing to date.

\subsection{SVO Traversal Algorithms}
Any implementation of the animation method introduced in this article will
as a matter of necessity have to employ an algorithm for the traversal of sparse
voxel octree models in order to fully facilitate the underlying ray tracing
procedure. The animation method is generally agnostic as to the underlying
traversal algorithm, but a brief introduction to the field is appropriate, and
will be presented in the following.

The earliest method found in the literature for octree traversal along a ray was
authored by Andrew Glassner in 1984 \cite{glassner84}. The paper reports that
over 95 percent of the total rendering time may be spent on ray-object
intersection calculations. Hence, there is a huge potential for performance gain
by optimising this process. Glassner then suggests sorting objects in the scene
into an octree and presents an algorithm for traversal of such an octree.
Another method was introduced by Marc Levoy in 1990 \cite{levoy90}. In the
paper, he introduces two different methods for enhancing the performance of
volumetric data ray tracing. The first of his methods is relevant in that
octrees are employed to encode spatial coherence in the data.

A number of subsequent attempts at improving the performance of octree traversal
exist. They can generally be grouped into two main categories based on how they
solve the traversal problem: \textit{bottom-up} and \textit{top-down} schemes
\cite{revelles00}. The algorithm by Glassner, as well as other, similar schemes
are instances of bottom-up octree traversal algorithms \cite{samet89,samet90}.
The method by Levoy, and a host of other algorithms \cite{revelles00, agate89,
jansen86, cohen93, gargantini93, endl94, knoll06, knoll09} provide examples of a
top-down parametric traversal algorithms. From the number of publications alone,
it appears that top-down traversal algorithms are most popular in the research
field.

An efficient algorithm for octree traversal was presented by Revelles, Ureña,
and Lastra in 2000 \cite{revelles00}. They introduce a top-down parametric
method that is very well documented. The algorithm is presented as recursive,
but due to its simplicity has been shown to translate well into an iterative
method, which is desirable for parallelisation \cite{wilhelmsen12}. After
\cite{revelles00} was published, there seems to be few new algorithms that
contest its speed and simplicity. An algorithm based on the work by
\cite{gargantini93} was introduced in 2006 by \cite{knoll06} (and subsequently
improved upon by the same authors \cite{knoll09}) which may be more efficient in
some circumstances. However, in addition to not being as well-documented as
\cite{revelles00}, the algorithm is recursive and according to \cite{laine11}
does not readily translate to an efficient implementation on a GPU or in
hardware.

\subsection{Parallelising the Workload}
The algorithms discussed in the previous section are sequential, single-threaded
algorithms which do not deliberately exploit the highly parallelisable nature of
ray tracing. In order to achieve real-time performance, the software
implementation developed as part of this work is parallelised and accelerated
using a GPU by means of the Nvidia CUDA API. It is therefore relevant to briefly
review works that discuss advantages of using the parallel computing
capabilities of GPUs to distribute the workload.

In 2009, \cite{crassin09} proposed a new approach for rendering large volumetric
data sets by ray tracing on a GPU. The result is a system which achieves
interactive to real-time performance while rendering several billion voxels. The
method takes care to avoid using a stack---and therefore no recursion---in order
to increase GPU optimisation. Mipmapping is utilised as a LoD-technique in order
to hide visual noise. The algorithm supports on-the-fly loading of data chunks
from CPU memory to GPU memory whenever the ray tracer encounters missing data,
which means that model size is not bounded by GPU memory.

Laine and Karras \cite{laine11} use the Nvidia CUDA API to exploit the
general-purpose capabilities of GPUs to efficiently trace static SVO models in
parallel. Alongside their ray tracer implementation, a compact SVO memory
structure is introduced. A simplified variant of this memory structure is
employed in the authors' software implementation.

Gobbetti and Marton \cite{gobbetti05} demonstrate the rendering of very large
surface models using \textit{out-of-core} data management, meaning that the
approach supports data sets too large to fit in working memory. In their paper,
they use hardware acceleration in the form of a GPU to parallelise the workload
of rendering the data sets, while natively supporting different levels of
detail. A similar method could perhaps be utilised in ray tracing hardware to
support very large, highly detailed models.

\subsection{Other Works of Significance}
As part of his master's thesis, the corresponding author of this article worked
towards a hardware implementation of an SVO ray tracer with support for
animation. Knowing that the animation method would be presented in this article
at a later stage, Espe \cite{espe19} treated the method as previous work in his
thesis. In his work, he focused on the implementation of the SVO traversal
algorithm before considering animation, and time constraints lead to the
animation of SVO models in hardware never being fully functional. As such, a
hardware implementation of the method to be presented in this article is still
an open problem.

Another work of interest is \cite{wilhelmsen12}, in which a hardware ray tracer
for static SVO models is implemented. The choice of algorithms closely match the
algorithms selected for the software implementation this article, and the
feasibility of a hardware implementation of these is demonstrated.

\section{The Method}\label{sec:solution}
Rigid-body animation can be achieved by modelling the scene as a system of rigid
bodies transformed relative to each other. Each individual body in this system
may be regarded as a static model with an associated transform that varies with
time. The models that make up the scene can be static models of any kind, for
instance SVO models. This means that rigid-body animation in SVO ray tracing may
be achieved by simply treating the scene as a set of independent SVOs, in which
each SVO is a static model with a corresponding transform. The process of
animation is accordingly reduced to modifying these transforms in a timely
manner. The internal data of each SVO may remain unmodified for the duration of
the animation.

The question naturally raised at this stage is how the associated transforms
could be applied to SVO models, and what significance this would have for the
SVO traversal algorithm. Although it would be the traditional approach for
polygonal models, it is not feasible to apply the transformation to every data
point contained in an SVO model during the rendering stage. This has been shown
in \cite{bautembach11} to destroy the hierarchical structure of the SVO, and
thus prohibit efficient traversal.

\begin{figure}[!t]
    \centering
    \subfloat[]{%
    \begin{tikzpicture}
        \begin{scope}[local bounding box=svo1s]
            \node[svobb, name path global=svo1-svobb] (svobb) {};
            \node[svo] (svo) at (svobb.south west) {};
            \node at (svo) {SVO$_A$};
        \end{scope}
        \begin{scope}[shift={($(svo1s.east)+(2.5, 0)$)}, rotate=45]
            \node[svobb, name path global=svo2-svobb] (svobb) {};
            \node[svo] (svo) at (svobb.south west) {};
            \node at (svo) {SVO$_B$};
        \end{scope}

        \draw[->, semithick, name path=ray] (-1.2, -1) node (start) {} --
            ++(15:6.7) node[above left, yshift=2pt] {$\vec{R}$};

        \path[name intersections={of=ray and svo1-svobb}];
        \node[circle, fill, inner sep=1.5pt] at (intersection-1) {};
        \node[below, yshift=-2pt] at (intersection-1) {$t_{enter}$};
        \node[circle, fill, inner sep=1.5pt] at (intersection-2) {};
        \node[below right, yshift=0pt] at (intersection-2) {$t_{exit}$};

        \path[name intersections={of=ray and svo2-svobb}];
        \node[circle, fill, inner sep=1.5pt] at (intersection-1) {};
        \node[above left, yshift=0pt] at (intersection-1) {$t_{enter}$};
        \node[circle, fill, inner sep=1.5pt] at (intersection-2) {};
        \node[below right, yshift=1pt, xshift=2pt] at (intersection-2) {$t_{exit}$};
    \end{tikzpicture}
    \label{fig:svo_ray_transform_1}}

    \subfloat[]{%
    \begin{tikzpicture}
        \begin{scope}[local bounding box=svo1s]
            \node[svobb, name path global=svo1-svobb] (svobb) {};
            \node[svo] (svo) at (svobb.south west) {};
            \node at (svo) {SVO$_A$};
        \end{scope}
        \begin{scope}[shift={($(svo1s.east)+(2.5, 0)$)}]
            \node[svobb, name path global=svo2-svobb] (svobb) {};
            \node[svo] (svo) at (svobb.south west) {};
            \node at (svo) {SVO$_B$};
        \end{scope}

        \draw[->, semithick, name path=ray1] (-1.2, -1) node (start1) {} --
            ++(15:3) node[above, yshift=2pt] {$\vec{R}'$};
        \draw[->, semithick, name path=ray2] (2.5, 1) node (start2) {} --
            ++(-30:3.2) node[below, yshift=-2pt] {$\vec{R}'$};

        \path[name intersections={of=ray1 and svo1-svobb}];
        \node[circle, fill, inner sep=1.5pt] at (intersection-1) {};
        \node[below, yshift=-2pt] at (intersection-1) {$t_{enter}$};
        \node[circle, fill, inner sep=1.5pt] at (intersection-2) {};
        \node[below right, yshift=0pt] at (intersection-2) {$t_{exit}$};

        \path[name intersections={of=ray2 and svo2-svobb}];
        \node[circle, fill, inner sep=1.5pt] at (intersection-1) {};
        \node[above, yshift=8pt] at (intersection-1) {$t_{enter}$};
        \node[circle, fill, inner sep=1.5pt] at (intersection-2) {};
        \node[above right, yshift=-2pt] at (intersection-2) {$t_{exit}$};
    \end{tikzpicture}
    \label{fig:svo_ray_transform_2}}
    \caption{Ray transformation to facilitate animation. The ray is shown in
    world space \protect\subref{fig:svo_ray_transform_1}, and in the local space
    of each SVO after the transformation
    \protect\subref{fig:svo_ray_transform_2}.}
    \label{fig:svo_ray_transform}
\end{figure}
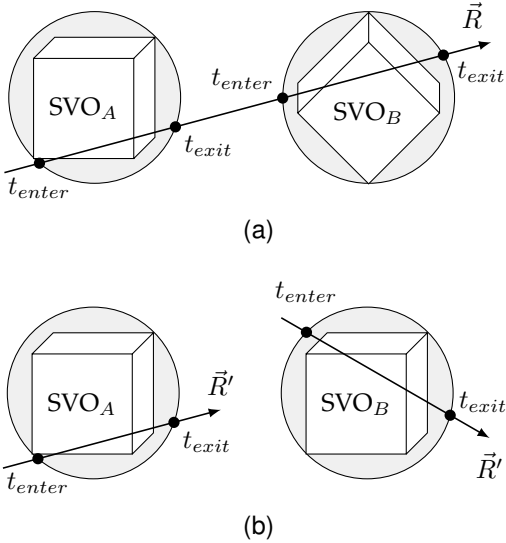

The novel approach presented here is to simply invert the problem. Instead of
transforming the model data during or after traversal, one may perform an
inverse transformation on the ray before traversal begins. In other words, each
ray in the ray tracing process will be transformed from world space to the local
co-ordinate system of the animated SVOs that are to be traced. The process is
illustrated by \Cref{fig:svo_ray_transform}, in which a ray is shown entering
the boundaries of two SVO models with different transforms. As the figure
highlights, the ray itself is transformed inversely in order to facilitate
animation of the models.

\subsection{Mathematical Formulation}
If the animation method is to be implemented in software or hardware, a
mathematical formulation for the ray transformation is necessary. For this
purpose, a ray is defined parametrically as
\begin{equation}
    \vec{R}(t; \; \mathbf{r}_o, \mathbf{r}_d) = \mathbf{r}_o + t \mathbf{r}_d \
    , \quad t \geq 0 \ , \label{eq:ray_def}
\end{equation}
where $\mathbf{r}_o$ is the ray origin and $\mathbf{r}_d$ is the ray direction.
A ray on this form will in the following be denoted by the arrow symbol. Given
the definition in \Cref{eq:ray_def}, a transformation $T$ from the ray $\vec{R}$
in world co-ordinates to the ray $\vec{R}'$ in the SVO's local co-ordinates must
be derived. The desired transformation should behave as shown in
\Cref{eq:ray_transform}.
\begin{equation}
    \vec{R}'(t; \; \mathbf{r}_o', \mathbf{r}_d') = T\left[\vec{R}(t;
    \; \mathbf{r}_o, \mathbf{r}_d)\right] \label{eq:ray_transform}
\end{equation}

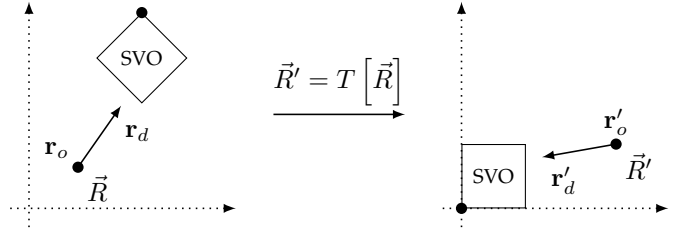
\begin{figure}[!t]
    \centering
    \begin{tikzpicture}

        \begin{scope}[shift={(1.5, 2)}, rotate=225]
            \node[square, scale=0.8] (svo1) {};
            \node[scale=0.8] at (svo1) {SVO};
            \node[circle, fill, inner sep=1.5pt] at (svo1.south west) {};
            \draw[->, semithick] ($(svo1.north east)+(1.2, 0)$)
            node[circle, fill, inner sep=1.5pt] {}
            node[right, yshift=-8pt] {$\vec{R}$}
            node[above left] {$\mathbf{r}_o$} -- ++(190:1)
            node[below right, yshift=-4pt, xshift=-2pt] {$\mathbf{r}_d$};
        \end{scope}
        \begin{scope}[local bounding box=svo2s, shift={(5.75, 0)}]
            \node[square, scale=0.8, anchor=south west] (svo2) {};
            \node[scale=0.8] at (svo2) {SVO};
            \node[circle, fill, inner sep=1.5pt] at (svo2.south west) {};
            \draw[->, semithick] ($(svo2.north east)+(1.2, 0)$)
            node[circle, fill, inner sep=1.5pt] {}
            node[below right] {$\vec{R}'$}
            node[above] {$\mathbf{r}_o'$} --
            ++(190:1) node[below right] {$\mathbf{r}_d'$};
        \end{scope}

        \draw[->, dotted, semithick] (0, -0.25) -- ++(0, 3);
        \draw[->, dotted, semithick] (-0.25, 0) -- ++(3, 0);

        \draw[->, dotted, semithick] ($(svo2.south west)-(0, 0.25)$) -- ++(0, 3);
        \draw[->, dotted, semithick] ($(svo2.south west)-(0.25, 0)$) -- ++(3, 0);

        \draw[->, semithick] (3.25, 1.25) -- ($(svo2.south west)+(-0.75, 1.25)$)
        node[above, midway] {$\vec{R}' = T\left[\vec{R}\right]$};

    \end{tikzpicture}
    \caption{The co-ordinate system transformation of a ray in world space to
    local space.}
    \label{fig:ray_transform_math}
\end{figure}

A graphical representation of the desired transformation is shown in
\Cref{fig:ray_transform_math}. By initially only allowing the SVO to be rotated
and translated, the mathematical derivation becomes straightforward. The
transformation is affine and results in a translation and rotation of the ray,
which means that formulating $T$ mathematically is simply a matter of
deriving two matrices with which to multiply the constituent vectors
$\mathbf{r}_o$ and $\mathbf{r}_d$ of the ray $\vec{R}$. Given the rotation
matrix $\mathbf{M}_R$ and the translation matrix $\mathbf{M}_T$ of the SVO, the
transformation function $T$ can be formulated as described in the following.
Since it makes no sense translating a directional vector, it should be
self-evident that the ray direction may only be influenced by the rotation of
the SVO. The ray origin, however, is affected by both rotation and translation.

The resulting definition of $T$ is shown in \Cref{eq:ray_transform_def}. The ray
direction is determined by simply premultiplying it with the inverse rotation of
the SVO. The ray origin is initially translated so that the origin of its
co-ordinate system is at the origin of the octree. Then the vector is rotated
around the SVO origin by the same inverse rotation as employed for the ray
direction.
\begin{equation}\label{eq:ray_transform_def}
    \begin{aligned}
        T : \; &\vec{R}(t; \; \mathbf{r}_o, \mathbf{r}_d) \mapsto
        \vec{R}'(t; \; \mathbf{r}_o', \mathbf{r}_d') \\
        &\text{such that } \begin{cases}
        \mathbf{r}_d' = \mathbf{M}^{-1}_R \mathbf{r}_d \\
        \mathbf{r}_o' = \mathbf{M}^{-1}_R \mathbf{M}^{-1}_T \mathbf{r}_o
        \end{cases}
    \end{aligned}
\end{equation}

Note that the inverse of translation and rotation matrices are quite simple to
attain in both software and hardware. The inverse of a translation matrix is
computed by negating its co-ordinates, and the inverse of a rotation matrix is
its transpose.

\subsection{Extending the Method to Allow Anisotropic Scaling}
The transformation so far only accounts for translation and rotation of SVO
models. Whereas these two affine transforms are the only ones strictly required
to provide the functionality of rigid-body animation, animation of model size is
often crucial element of many animation sequences.

The most convenient way of supporting scaling in the scheme which has been
presented is to implement the support at the traversal stage of the ray tracing
process. Fortunately, many SVO traversal algorithms are designed to allow the
tuning of octree dimensions. By employing this directly in the animation
process, the method presented above may remain simple, and only take the
rotation and translation into account. As an example, in the traversal algorithm
presented in \cite{revelles00}, the dimensions of the octree are defined
initially as a set of algorithm parameters. This approach does, however, make
the support for anisotropic scaling dependent on the choice of traversal
algorithm.

\section{Optimisations}\label{sec:optimisations}
A number measures may be taken in order to further improve the efficiency of the
method. The most effective optimisations explored in the authors' software
implementation will be discussed in the following as a supplement to the
animation technique.

\subsection{Bounding-Sphere Tests}
As a result of the sheer number of intersection tests, the most computationally
heavy stage of the ray tracing process is the traversal of the SVO
\cite{glassner84}. It is therefore desirable to solely traverse SVOs that can
potentially lead to a ray hit. For instance, consider a situation where the
origin of the ray is outside the bounds of the SVO model, and the the ray points
in the opposite direction to that of the octree. In such circumstances, the SVO
can safely be eliminated from the process, as it is impossible for the ray to
ever hit it.

There are many criteria one may use to determine which octrees that will never
be hit by a given ray, some more efficient in implementation than others. A
method that will be explored in the following is to perform an intersection test
between the ray and the bounding sphere of the SVO. This intersection test is
fast and can be solved analytically. An explicit form of the equation to be
evaluated is presented in \Cref{eq:sphere_int}, where $\mathbf{r}_o$ and
$\mathbf{r}_d$ are defined as earlier and the point $\mathbf{s}_o$ is the centre
of the sphere. If the resulting value of $d$ is real and less than the sphere
radius, the ray intersects the sphere.
\begin{equation}
    \mathbf{l} = \mathbf{s}_c - \mathbf{r}_o \ ,
    \qquad d = \sqrt{\mathbf{l} \cdot \mathbf{l} - \left(\mathbf{l} \cdot
    \mathbf{r}_d\right)^2} \label{eq:sphere_int}
\end{equation}

The rationale behind choosing a bounding sphere instead of a bounding box to
represent the bounds of an octree model is that the bounding sphere is invariant
under rotation of the corresponding SVO. The only properties to consider when
constructing the sphere is the position of its centre and its radius, both of
which are readily obtainable from the SVO; indeed, they are given directly by
its translation and scale. This means that the ray-sphere intersection test
remains simple even though the underlying SVO may have an arbitrary orientation.
A drawback of using bounding spheres instead of bounding boxes is that it will
on occasion lead to false positives. Situations may arise where an octree model
is processed and traversed even though the ray does not intersect with the SVO.

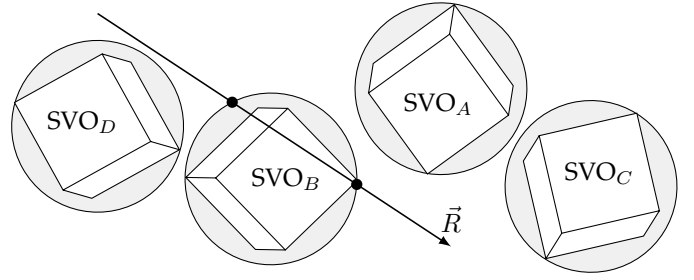
\begin{figure}[!t]
    \centering
    \begin{tikzpicture}
        \begin{scope}[local bounding box=svo1s, rotate=-61]
            \node[svobb] (svobb) {};
            \node[svo] (svo) at (svobb.south west) {};
            \node at (svo) {SVO$_D$};
        \end{scope}
        \begin{scope}[local bounding box=svo2s,
            shift={($(svo1s.east)+(1, -0.7)$)}, rotate=134]
            \node[svobb, name path global=svo2-svobb] (svobb) {};
            \node[svo] (svo) at (svobb.south west) {};
            \node at (svo) {SVO$_B$};
        \end{scope}
        \begin{scope}[local bounding box=svo3s,
            shift={($(svo2s.east)+(1, 1.2)$)}, rotate=36]
            \node[svobb] (svobb) {};
            \node[svo] (svo) at (svobb.south west) {};
            \node at (svo) {SVO$_A$};
        \end{scope}
        \begin{scope}[local bounding box=svo4s,
            shift={($(svo3s.east)+(0.7, -1.3)$)}, rotate=193]
            \node[svobb] (svobb) {};
            \node[svo] (svo) at (svobb.south west) {};
            \node at (svo) {SVO$_C$};
        \end{scope}

        \draw[->, semithick, name path=ray] (0, 1.5) --
            (4.7, -1.6) node[above, yshift=2pt] {$\vec{R}$};

        \path[name intersections={of=ray and svo2-svobb}];
        \node[circle, fill, inner sep=1.5pt] at (intersection-1) {};
        \node[circle, fill, inner sep=1.5pt] at (intersection-2) {};
    \end{tikzpicture}
    \caption{Using the bounding sphere of an SVO to avoid traversing octrees
    that will be missed. $\mathrm{SVO}_B$ is the only octree that will be traversed in
    this case.}
    \label{fig:svo_bounding_sphere}
\end{figure}

In \Cref{fig:svo_bounding_sphere} the principle is illustrated. A scene
consisting of four animated SVOs is shown, where bounding spheres are utilised
to determine which octree models that will be missed by the ray. In this case,
only SVO$_B$ will be traversed, as the bounding sphere of the other three
octrees in the scene do not intersect the ray.

\subsection{Depth Sorting}
The use of bounding spheres also lends itself to another optimisation which will
be elaborated in the following. The general idea is that SVO models may be
traced in a front-to-back order, sorted by the distance along the ray for each
intersected bounding sphere. Once traversal of an SVO model results in a ray hit
that is closer than the bounding sphere of the next SVO to be traced, the ray
tracing process may be stopped early, as no subsequent object can lie in front
of the current hit.

\begin{figure}[!t]
    \centering
    \begin{tikzpicture}
        \begin{scope}[local bounding box=svo1s]
            \node[svobb, scale=0.8] (svobb) {};
            \node[svo, scale=0.8] (svo) at (svobb.south west) {};
            \node[scale=0.8] at (svo) {SVO$_D$};
        \end{scope}
        \begin{scope}[local bounding box=svo2s,
            shift={($(svo1s.east)+(0.8, 1)$)}, rotate=110]
            \node[svobb, scale=0.8] (svobb) {};
            \node[svo, scale=0.8] (svo) at (svobb.south west) {};
            \node[scale=0.8] at (svo) {SVO$_B$};
        \end{scope}
        \begin{scope}[local bounding box=svo3s,
            shift={($(svo2s.east)+(0.4, -1.4)$)}, rotate=205]
            \node[svobb, scale=0.8] (svobb) {};
            \node[svo, scale=0.8] (svo) at (svobb.south west) {};
            \node[scale=0.8] at (svo) {SVO$_A$};
        \end{scope}
        \begin{scope}[local bounding box=svo4s,
            shift={($(svo3s.east)+(1, 1)$)}, rotate=-47]
            \node[svobb, scale=0.8] (svobb) {};
            \node[svo, scale=0.8] (svo) at (svobb.south west) {};
            \node[scale=0.8] at (svo) {SVO$_C$};
        \end{scope}

        \draw[->, semithick] (-1.5, 0) -- ++(3.5:4.9) node (pt) {}
        -- ++(3.5:3.5) node[below, yshift=-2pt] {$\vec{R}$};

        \node[circle, fill, inner sep=1pt] at (pt) {};
        \node[above right, yshift=2pt] at (pt) {$t_{hit}$};
    \end{tikzpicture}
    \caption{Tracing a sorted list of SVOs. The process may end after
    $\mathrm{SVO}_A$ has been traversed, and disregard $\mathrm{SVO}_C$.}
    \label{fig:svo_sorted}
\end{figure}
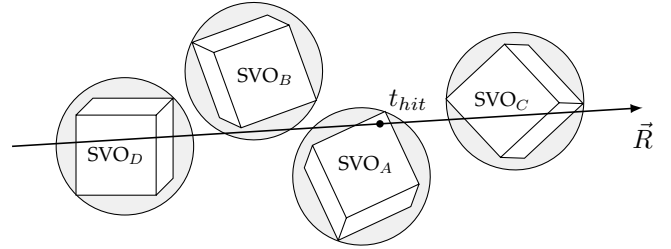

An example of the depth sorting technique is illustrated in
\Cref{fig:svo_sorted}, where a scene consisting of four animated SVOs is shown.
The tracing is performed in a sorted manner, the order of traversal determined
by the distance to the bounding sphere centre. In this example the order of
increasing depth will be $\{D, B, A, C\}$. The figure shows a situation where
SVO$_D$ is traversed without a hit, SVO$_B$ is traversed as a false positive,
SVO$_A$ results in a trace hit, and SVO$_C$ is not traversed. The ray tracing
process is terminated after the third SVO model since it results in a ray hit,
and the distance along the ray of this hit is closer than the distance to the
boundary of the next octree that would be traversed. In other words, it is
mathematically impossible that traversal of SVO$_C$ will yield a ray hit closer
than the hit produced by traversal of the third octree.

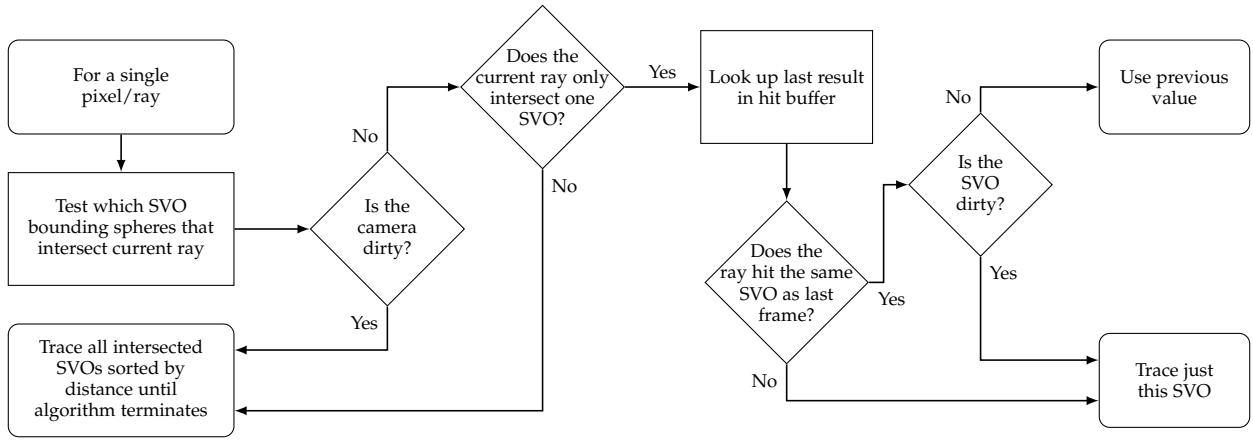
\begin{figure*}[!t]
    \centering
    \begin{tikzpicture}[node distance=0, font=\scriptsize]
        \node[square, rounded corners=4pt, minimum height=1.25cm,
            minimum width=3cm, align=center] (init) {For a single \\ pixel/ray};

        \node[square, minimum width=3cm, minimum height=1.5cm, below=0.5 of init,
            align=center] (act1)
            {Test which SVO \\ bounding spheres that \\ intersect current ray};

        \node[diamond, draw=black, right=1 of act1, align=center] (test1)
            {Is the \\ camera \\ dirty?};

        \node[diamond, draw=black, right=3 of init, align=center,
            inner sep=-1ex] (test2) {Does the \\ current ray only \\
            intersect one \\ SVO?};

        \node[square, rounded corners=4pt, minimum width=3cm,
            minimum height=1.5cm, below=0.5 of act1, align=center] (term1)
            {Trace all intersected \\ SVOs sorted by \\ distance until \\ algorithm
            terminates};

        \node[square, minimum width=2cm, minimum height=1.5cm, right=1 of test2,
            align=center] (act2) {Look up last result \\ in hit buffer};

        \node[diamond, draw=black, below=0.75 of act2, align=center, inner
            sep=-1ex] (test3) {Does the \\ ray hit the same \\ SVO as last \\
            frame?};

        \node[square, rounded corners=4pt, minimum width=2cm,
            minimum height=1.25cm, right=3 of act2, align=center] (term2)
            {Use previous \\ value};

        \node[square, rounded corners=4pt, minimum width=2cm,
            minimum height=1.25cm, align=center]
            (term3) at (term2|-term1) {Trace just \\ this SVO};

        \node[diamond, draw=black, align=center] (test4) at
            ($(test3)!0.5!(term2)$) {Is the \\ SVO \\ dirty?};

        \draw[->, semithick] (init) -- (act1);
        \draw[->, semithick] (act1) -- (test1);
        \draw[->, semithick] (test1.south) node[below left] {Yes} |- (term1.15);
        \draw[->, semithick] (test1.north) node[above left] {No} |- (test2);
        \draw[->, semithick] (test2.south) node[below right] {No} |- (term1.345);
        \draw[->, semithick] (test2.east) -- node[above, midway] {Yes} (act2);
        \draw[->, semithick] (act2.south) -- (test3);
        \draw[->, semithick] let \p1=(test3), \p2=(test4) in (test3.east)
            node[below right] {Yes} -- ({0.5*(\x1 + \x2)}, \y1) |- (test4);
        \draw[->, semithick] (test3.south) node[below left] {No} |- (term3.195);
        \draw[->, semithick] (test4.north) node[above left] {No} |- (term2);
        \draw[->, semithick] (test4.south) node[below right] {Yes} |- (term3.165);
    \end{tikzpicture}
    \caption{The hit buffer algorithm.}
    \label{fig:hit_buffer_algo}
\end{figure*}

\subsection{Hit Buffer Algorithm}
A simple buffering mechanism was developed as part of the work. The
buffer---termed \textit{hit buffer object} (HBO)---stores the ray tracing result
for each pixel in the last rendered frame. The idea is that if, for a given
pixel, the scene is unchanged \textit{enough} since the last frame, the
traversal of SVOs for this pixel may be streamlined, or even skipped, in which
case the value from the last frame is reused. The algorithm is illustrated as a
flow chart in \Cref{fig:hit_buffer_algo}.

Stored for each pixel in the HBO data structure are: the colour, the normal, the
$t$-value of the hit along the ray (i.e. the depth, which means that the HBO
also functions as a depth buffer), a unique identifier of the SVO that was hit,
and the nature of the hit. The last field provides information for the hit
buffer algorithm, such as whether nothing was hit, or whether the ray passed
through a single or multiple bounding spheres before arriving at the hit point.

The HBO is used in conjunction with a set of state variables to determine if the
value of a pixel is unchanged between frames. Each SVO in the scene, as well as
the camera, has a flag that specifies if the object has moved since the last
frame (if it is \textit{dirty}). The application can then look up the hit buffer
data for the current pixel and, if both the camera and the SVO object hit by the
ray the last frame are unchanged, simply use the last value and avoid tracing
the SVO again. It is also required that the ray did not pass through multiple
bounding spheres before the hit for this optimisation to take place. The reason
for this requirement is that if the ray passes through other SVOs, these might
have changed in the meantime, and there is no way of determining if the ray
would hit data in these SVOs without traversing them again.

\section{Evaluation}\label{sec:evaluation}
A software demonstration of the method was prepared for this article. The
implementation was written in the C programming language and employs the APIs
Nvidia CUDA and OpenGL---the former for parallelising the ray tracing process,
and the latter for easing the process of displaying the result. Three different
GPUs were available for testing: the \textit{Nvidia Quadro T1000} \cite{t1000},
the \textit{Nvidia GeForce GTX 680} \cite{gtx680}, and the \textit{Nvidia Quadro
P5000} \cite{p5000}. The most relevant specifications for these are listed in
\Cref{tab:gpus}.

\begin{table}
    \renewcommand{\arraystretch}{1.3}
    \caption{GPUs employed in the test setup. All GPUs are from Nvidia.}
    \label{tab:gpus}
    \centering
    \begin{tabular}{lSS[table-format=4]S[table-format=4]r}
        \toprule
        {GPU name} & {Type} & {Year} & {CUDA cores} & {Base freq.}\\
        \midrule
        Quadro T1000 & {Mobile} & 2019 & 768 & 1395 MHz \\
        GeForce GTX 680 & {Desktop} & 2012 & 1536 & 1006 MHz \\
        Quadro P5000 & {Desktop} & 2016 & 2560 & 1164 MHz \\
        \bottomrule
    \end{tabular}
\end{table}

In addition to the animation method and the optimisations presented in this
article, the software implementation employs the work of \cite{revelles00} and
\cite{laine11}. In \cite{revelles00}, an efficient method for octree traversal
is described. This traversal algorithm was chosen since it is simple and fast,
and also improves upon the performance of earlier algorithms. Other, more
efficient algorithms exist (such as \cite{knoll06}), but since the focus of this
article is to showcase an animation method, greater emphasis is placed on the
simplicity of the algorithm rather than its pure speed and efficiency.
Introduced by \cite{laine11} is a data structure scheme for storing SVO data. A
simplified version of this structure was used for the SVO models in the authors'
implementation.

\begin{figure*}[!t]
    \centering
    \subfloat[]{\includegraphics[width=2.2in]{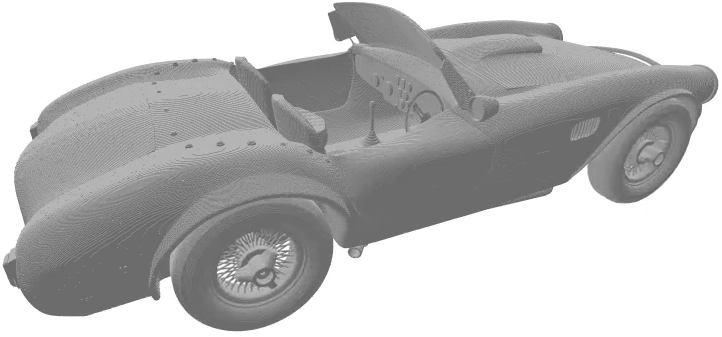}
    \label{fig:car_snap_1}}
    \hfill
    \subfloat[]{\includegraphics[width=2.2in]{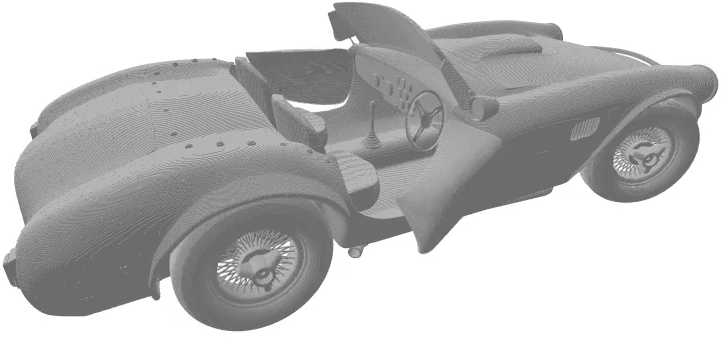}
    \label{fig:car_snap_2}}
    \hfill
    \subfloat[]{\includegraphics[width=2.2in]{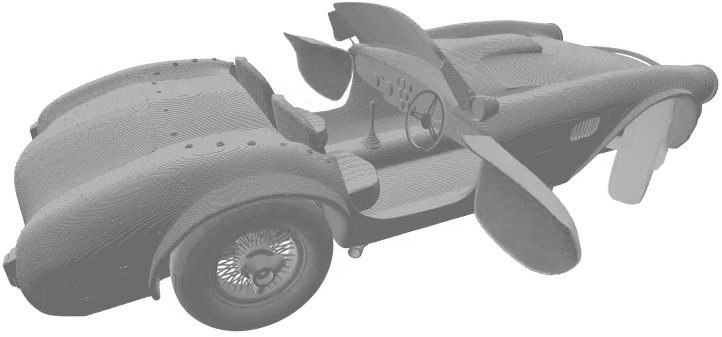}
    \label{fig:car_snap_3}}
    \caption{Three snapshots of an animated car rendered by the software
    implementation. The wheels are rolling, the doors open and close, and the
    steering wheel and front wheels turn left and right.}
    \label{fig:animation}
\end{figure*}

The software implementation was employed to render an animated model of a car,
illustrated by snapshots in \Cref{fig:animation}. The source model was obtained
from \cite{accobra}, and further processed in two steps to create the SVO
models. First, a raw voxel model was generated by rasterising the \texttt{.obj}
polygonal model using the \textit{Binvox} program \cite{binvox}. Secondly, a
custom program was developed in order to reduce the raw voxel model to an SVO
model on the format introduced by \cite{laine11}. The result was an SVO data
structure with an hierarchical depth of 11, and a total of $2048^3$ data points.
In order to establish a comparative basis, both static and animated versions of
the scene were rendered. In the animated scene, the car body, the wheels, the
doors, and the steering wheel are each realised as separate SVO models.

\subsection{Performance}
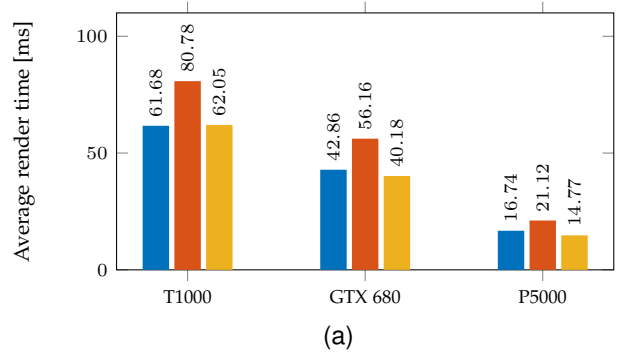
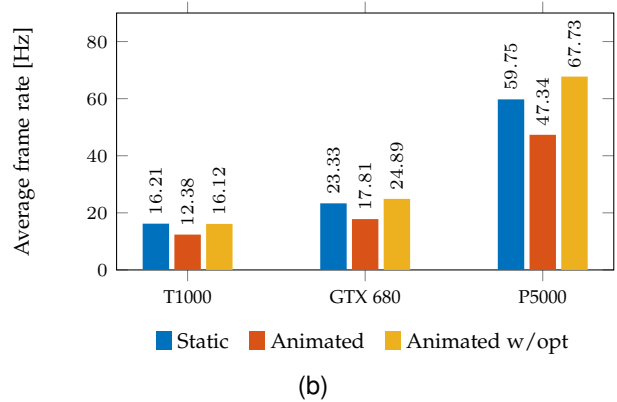
\begin{figure}[!t]
    \subfloat[]{\begin{tikzpicture}
        \begin{axis}[barplot, symbolic x coords={T1000, GTX 680, P5000},
            ylabel={Average render time [ms]}, ymin=0, ymax=110, title={Recorded
            performance of the software implementation \\ using three different
            GPUs}]
            \addplot[fill=matlab1, draw=none] coordinates {
                (T1000, 61.68)
                (GTX 680, 42.86)
                (P5000, 16.74)
            };
            \addplot[fill=matlab2, draw=none] coordinates {
                (T1000, 80.78)
                (GTX 680, 56.16)
                (P5000, 21.12)
            };
            \addplot[fill=matlab3, draw=none] coordinates {
                (T1000, 62.05)
                (GTX 680, 40.18)
                (P5000, 14.77)
            };
        \end{axis}
    \end{tikzpicture}
    \label{fig:render_time}}

    \subfloat[]{\begin{tikzpicture}
        \begin{axis}[barplot, symbolic x coords={T1000, GTX 680, P5000},
            ylabel={Average frame rate [Hz]}, ymin=0, ymax=90,
            legend style={at={(0.5,-0.2)}, draw=none, font=\footnotesize, anchor=north,
            legend columns=-1, column sep=0.2cm, /tikz/every odd column/.append
            style={column sep=0cm}}, legend image code/.code={%
                \draw[#1, draw=none] (0pt, -3pt) rectangle (6pt, 4pt);
            }]
            \addplot[fill=matlab1, draw=none] coordinates {
                (T1000, 16.21)
                (GTX 680, 23.33)
                (P5000, 59.75)
            };
            \addplot[fill=matlab2, draw=none] coordinates {
                (T1000, 12.38)
                (GTX 680, 17.81)
                (P5000, 47.34)
            };
            \addplot[fill=matlab3, draw=none] coordinates {
                (T1000, 16.12)
                (GTX 680, 24.89)
                (P5000, 67.73)
            };
            \legend{Static, Animated, Animated w/opt}
        \end{axis}
    \end{tikzpicture}
    \label{fig:frame_rate}}
    \caption{Average render time \protect\subref{fig:render_time} and frame rate
    \protect\subref{fig:frame_rate} for different model types rendered at
    1280$\times$768 resolution. Three different GPUs are employed to render a
    static model (blue), an animated model (red), and an animated model with
    optimisations enabled (yellow). The data are averaged over 60 seconds.}
    \label{fig:num_data}
\end{figure}

Shown in \Cref{fig:num_data} are the average render times and frame rates for
three different scenes rendered with the different GPUs of the test setup.
Running at a resolution of $1280 \times 768$, the implementation gathered data
over 60 seconds for each test case. The first scene featured a static model of a
car realised as a single SVO model analogous to the first snapshot shown in
\Cref{fig:car_snap_1}. Since animation was not enabled for the model, the
situation illustrates the performance of a pure implementation of the SVO
traversal algorithm, without any of the overhead associated with animation. The
second scene contained an animated car with each part of the body realised as
separate SVOs. It represents an implementation of the solution proposed in this
article without any of the optimisation techniques introduced in
\Cref{sec:optimisations}. The third scene is the animated scene rendered with
the bounding sphere optimisation techniques and the hit buffer mechanism
enabled.

\begin{figure}[!t]
    \centering
    \includegraphics[width=.4\textwidth]{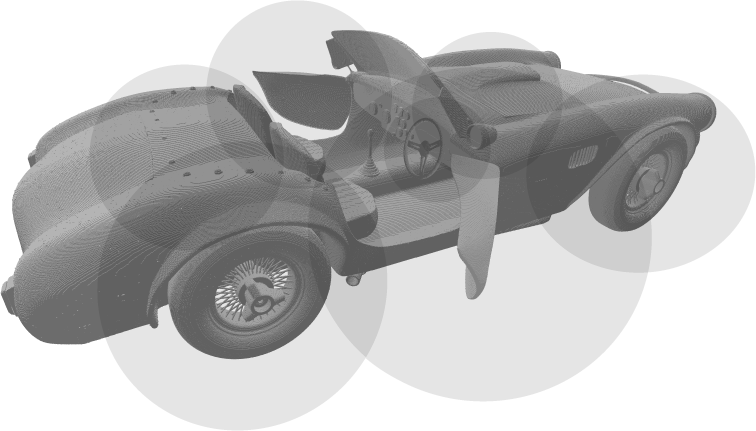}
    \caption{The animated car model with the bounding spheres for each SVO
    highlighted.}
    \label{fig:car_bounding_sphere}
\end{figure}

The last two scenes were visually identical and resulted in the animation
sequence shown in \Cref{fig:animation}. All three scenes were traced with a
static camera to ensure that the circumstances of the tests were equal and that
the performance could be compared. In addition, if a dynamic camera were to be
used the hit buffer algorithm enabled for the last scene would be rendered
ineffective. The bounding spheres used as part of the optimisation techniques
are shown highlighted in \Cref{fig:car_bounding_sphere} for illustrative
purposes.

At which point application performance can be classified as real-time is not
well-defined. Nonetheless, it can be argued in good faith that the performance
of the P5000 must clearly be classified as real-time computer graphics, as a
frame rate of \SI{67}{\hertz} is higher the refresh rate of most common computer
monitors. It is therefore shown that the animation method can deliver real-time
performance on current generation GPUs. For the T1000 GPU, the difference in raw
performance between the static model and the animated model with optimisations
enabled is quite small---the decrease in frame rate is \SI{0.6}{\percent}.
However, in the case of the GTX 680 and P5000 GPUs, the frame rate increases by
\SI{6.7}{\percent} and \SI{13.4}{\percent}, respectively. These numbers indicate
that most of the execution time is spent in the traversal section of the
software, not as a part of the animation process itself. In addition, it seems
that the optimisations may be more effective at higher frame rates, and will in
some situations improve upon the performance of the underlying SVO traversal
algorithm. The results lend credence to the claim that the method is suitable
for real-time animation of SVO data. Moreover, since the animation technique
does not introduce a noticeable overhead, it is expected that further
improvements in performance may be achievable with even more optimisation in the
traversal algorithm itself.

It should be noted that the results presented are gathered from a single
animated model. It is therefore not completely certain that the data fully
represent the general case. As part of future work, it would be advantageous to
run the implementation across a wide spectrum of different models and situations
in order to gather more data. This could be used to further investigate the
claim that the animation does not tax the performance noticeably, and help gain
understanding of how the method and optimisations responds to the rendering of
different data sets.

\subsection{Memory Requirements}
In terms of memory, the proposed solution does not represent a significant
increase in space, as only a few more bytes are required per SVO in order to
enable animation. Whereas a typical SVO model may be several megabytes in size
(the SVO for the car body shown in \Cref{fig:animation} has a size of
\SI{11}{\mega\byte}), a $4 \times 4$ homogeneous transform matrix only requires
\SI{64}{\byte} when using 32-bit floating-point numbers. As for memory
bandwidth, the proposed solution should perform very well. This is because the
only data that need to be updated between frames are the transform matrices, and
in some cases the bounding spheres and scales.

Even though each octree in theory only needs a single transform matrix to
specify its translation, rotation, and scale, it might be beneficial to store
the bounding sphere and scale separately in order to speed up execution.

\subsection{Suitability for Hardware Implementation}
The animation method itself is agnostic with regard to the underlying traversal
algorithm and the SVO data structure. Nonetheless, the algorithms chosen for the
software implementation have already been demonstrated to work in hardware
\cite{wilhelmsen12,espe19}. The animation logic itself is iterative and consists
almost exclusively of matrix multiplication which can readily be pipelined and
accelerated by dedicated circuits. The optimisations should also be relatively
straightforward to translate into hardware. The hit buffer algorithm will
require memory and simple Boolean logic, whereas the most complex part of the
ray-sphere intersection test will be a square root hardware module, as shown in
\Cref{eq:sphere_int}. Since a hardware implementation will be relatively simple,
a consequence is that the computations may be kept on-chip, allowing shorter
critical paths, and a higher core frequency.

In the discussion of memory requirements, the conclusion was drawn that the
proposed solution should not be limited by memory bandwidth since a very small
amount of data has to be updated in memory between frames. This also translates
well into direct improvements in a hardware setting. By limiting the memory
bandwidth, a hardware implementation may avoid performing accesses to external
memory altogether, further increasing performance. In addition, since the sparse
voxel octree data remain unchanged when animated, one or more levels of caching
may be employed in order to reduce the latency associated with memory accesses.
Lastly, the amount of data that describes the animation of an octree is very
small, which means that these transformation matrices may be stored directly in
very fast registers close to the core, further reducing memory access latency.

\subsection{Limitations}
The main limitation of the animation method is that it only supports rigid-body
animation, which will limit its applications. Effects such as deformation,
bending, and other features which are often needed to enable complex and
life-like animations are not supported. Consequently, the animation
functionality provided by the solution is mostly suited for stiff, mechanical
objects.

There may also be issues with the method's scalability, since each part of the
scene that is to be animated is required to be stored as a separate SVO with an
associated transform. A complex object with hundreds of moving parts would have
to be formulated as hundreds of independent SVOs, and every ray would have to be
intersection-tested with the bounding sphere of each SVO in the scene. Future
work may include exploring measures that can be taken to mitigate problems
stemming from these predicted scalability issues. An idea to this end is to sort
the entire scene into a larger octree so that only relevant SVOs would have to
be considered by the ray tracing algorithm.

\subsubsection{Limitations of the Optimisation Techniques}
There is a limit to the region of applicability for most optimisation
techniques. The hit buffer optimisation is limited in that becomes ineffective
once the camera moves. This is a consequence of the fact that once the camera is
altered in some fashion, virtually all rays have also necessarily changed since
the last frame. As such, the previous results stored in the hit buffer are no
longer applicable, and the entire scene must be retraced.

The bounding sphere optimisation techniques are still effective regardless of
camera movement, as they are optimisations in world space, not in screen space.
In other words, these optimisations are amendments to the ray tracing algorithm
itself, and not dependent on the camera, and thus should provide a performance
gain regardless of camera movement.

\section{Conclusion}
In this article, a method for animation of SVO models was introduced. The method
circumvents the need to rebuild the octree structure for each frame of an
animation sequence, and is therefore suited for real-time computer animation.
Being relatively simple to implement in software, the method is agnostic in
regards to the underlying octree traversal algorithm. It facilitates rigid-body
animation, meaning that rotation and translation of SVO models are supported.
Anisotropic scaling is in many cases also possible, depending on the choice of
traversal algorithm.

Using a set of optimisation techniques, it was shown that the general
performance of the animation method may be enhanced. Initially, bounding spheres
were introduced as a means to determine whether SVO models could be preemptively
excluded from the rendering process. It was also demonstrated that these
bounding spheres may be utilised for depth sorting, so that the object in the
scene can be traversed in the most efficient order possible. Lastly, a buffering
mechanism termed the hit buffer algorithm was presented, which takes advantage
of certain situations where parts of the scene are unchanged to reduce the
required processing load.

A software implementation of the animation method was introduced to permit a
direct performance comparison between the rendering of static and animated SVO
models. The software implementation was tested using three available GPUs on
consumer hardware, with results testifying that the rendering of an animated
model with optimisations enabled generally outperforms its static counterpart.
The software implementation demonstrates real-time performance while rendering
animated SVO models on a desktop computer, with the strongest GPU reporting an
average frame rate of about \SI{67}{\hertz}---a frame rate higher the refresh
rate of most common computer monitors.

The method is well-suited for implementation in hardware; since the method is
relatively simple, it should not pose serious implementation challenges beyond
those associated with a hardware implementation of the traversal algorithm
itself. On account of real-time performance already being demonstrated when
running the method in software, it is expected that an implementation in
application-specific hardware would be able to deliver real-time performance of
more complex models and scenes, and at larger resolutions. It will be
interesting to follow any further developments, as it has been predicted that
ray tracing may very well play a central role in the future of real-time
computer graphics \cite{caulfield18}. A hardware implementation of the method is
certainly something that could be explored as part of future work, perhaps
building on related works such as \cite{wilhelmsen12} or \cite{espe19}.

\bibliographystyle{IEEEtran}
\bibliography{IEEEabrv,paper}

\begin{IEEEbiography}[{
    \includegraphics[trim={0.5cm 0 0.5cm 0}, width=1in, height=1.25in, clip, keepaspectratio]
    {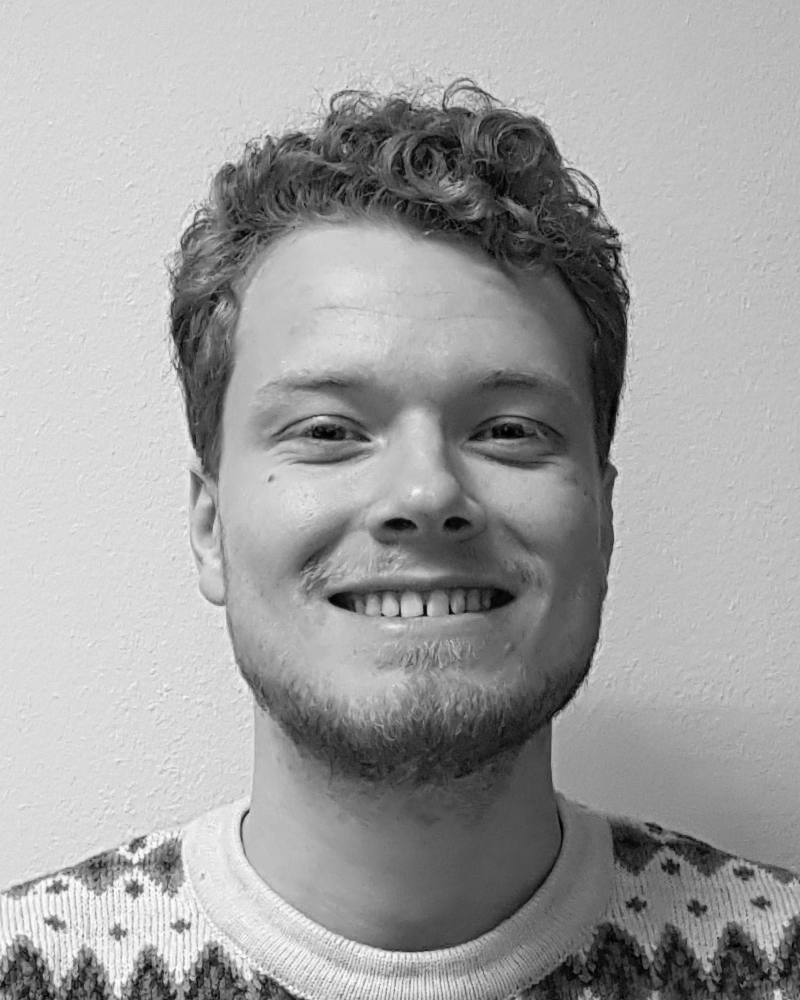}
    }]{Asbjørn Engmark Espe}
is a PhD candidate at the Department of Engineering Cybernetics, Norwegian
University of Science and Technology (NTNU). His research interests include
low-power and low-maintenance embedded systems, ray tracing, and wireless sensor
networks. Espe received his MS degree in cybernetics and robotics from NTNU in
2019. He is a Student Member of the IEEE. Contact him at asbjorn.e.espe@ntnu.no.
\end{IEEEbiography}

\begin{IEEEbiography}[{
    \includegraphics[width=1in, height=1.25in, clip, keepaspectratio]
    {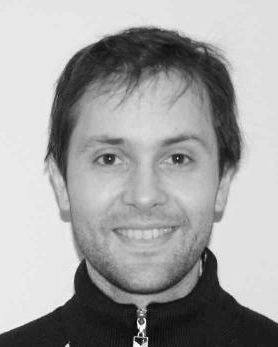}
    }]{Øystein Gjermundnes}
is an associate professor at the Department of Electronic Systems, Norwegian
University of Science and Technology (NTNU). His research interests include
hardware design and verification. Gjermundnes received his MS degree in
microelectronics from NTNU in 2002, and his PhD degree from NTNU in 2006.
Contact him at oystein.gjermundnes@ntnu.no.
\end{IEEEbiography}

\begin{IEEEbiography}[{
    \includegraphics[width=1in, height=1.25in, clip, keepaspectratio]
    {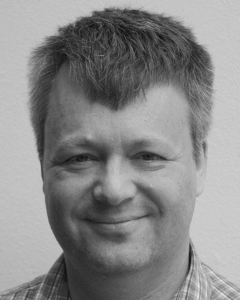}
    }]{Sverre Hendseth}
is an associate professor at the Department of Engineering Cybernetics,
Norwegian University of Science and Technology (NTNU). His research interests
include real-time systems, programming languages, and software engineering.
Hendseth received his MS degree in engineering cybernetics from the Norwegian
Institute of Technology (NTH) in 1987, and his Dr.Ing. degree from NTH in 1994.
He is a Member of the IEEE. Contact him at sverre.hendseth@ntnu.no.
\end{IEEEbiography}

\end{document}